\documentclass[11pt]{article}
\usepackage[utf8]{inputenc}
\usepackage[english]{babel}
\usepackage{amsmath}
\usepackage{parskip}
\usepackage{dsfont}
\usepackage{enumitem}
\usepackage{graphicx}
\usepackage{caption}
\usepackage{float}
\usepackage{amssymb}
\usepackage{cuted}
\usepackage{lipsum}
\usepackage{abstract}
\numberwithin{equation}{section}
\usepackage[left=1.5 cm, right=1.5 cm , top= 1.5 cm, bottom= 3 cm]{geometry}
\usepackage[backend=biber, style=phys]{biblatex}
\usepackage{hyperref}
\bibliography{Bibliography}

\usepackage{authblk}
\hypersetup{
    colorlinks=true,
    linkcolor=blue,
    filecolor=blue,  
    citecolor= blue,
    urlcolor=blue
    }
\date{}

\bibliography{Bibliography}

\title{Expanding CGL: The CGL double-adiabatic approximation in the Expanding Solar Wind}

\author[1,*]{Sebastián Echeverría-Veas}
\author[1,*]{Pablo S. Moya}
\author[2,3]{Marian Lazar}
\author[2,4]{Stefaan Poedts}
\author[5]{Felipe A. Asenjo}

\affil[1,]{ Departamento de Física, Facultad de Ciencias, Universidad de Chile, Las Palmeras 3425, Ñuñoa, Santiago 7800003, Chile; $^*$s.echeverria@ug.uchile.cl (S. E-V), pablo.moya@uchile.cl (P.S.M).}
\affil[2]{Centre for mathematical Plasma Astrophysics, Dept.\ of Mathematics, KU Leuven, Celestijnenlaan 200B, B-3001 Leuven, Belgium.}
\affil[3]{Institute for Theoretical Physics IV, Faculty for Physics and Astronomy, Ruhr University Bochum, D-44780 Bochum, Germany.}
\affil[4]{ Institute of Physics, University of Maria Curie-Skłodowska, ul.\ Radziszewskiego 10, 20-031 Lublin, Poland.}
\affil[5]{Facultad de Ingeniería y Ciencias, Universidad Adolfo Ibáñez, Santiago 7491169, Chile.}

\begin{document}

\maketitle

\begin{abstract}
\vspace{1cm}
Different \textit{in situ} satellite observations within 0.3 to 1 AU from the Sun reveal deviations in the thermodynamics of solar wind expansion. Specifically, these deviations challenge the applicability of the double adiabatic or CGL theory, indicating potential influences such as perpendicular heating and/or parallel cooling of ions. The study aims to investigate the plasma expansion phenomena using the Expanding Box Model (EBM) coupled with an ideal MHD description of the plasma. The primary objective is to understand the observed deviations from the CGL predictions, and how the expansion can affect the conservation of the adiabatic invariants, particularly focusing on the impact of transverse expansion on the CGL equations. To address the plasma expansion, we employed the Expanding Box Model (EBM) coupled with the ideal-MHD formalism used for CGL theory. This model provides a unique system of reference co-moving with the solar wind, allowing for the incorporation of transverse expansion into the double adiabatic equations. An ideal EBM-MHD model was studied within this formalism to develop a CGL-like theory in the Expanding Box (EB) frame. New equations were derived, explicitly accounting for transverse velocity gradients and modifying the conservation of double adiabatic predictions. Our results show the EBM modifies the conservation of the double adiabatic equations, through an explicit dependence on expanding parameters, magnetic field profiles, and velocity gradients, which have not been addressed in the literature. Solving the equations for different magnetic field profiles, we compute the evolution of anisotropy and plasma beta, which deviates from CGL predictions and empirical observations. This deviation is attributed to the plasma cooling effect induced by the Expanding Box Model (EBM). Results suggest that heating mechanisms play a crucial role in counteracting plasma cooling during expansion.
\end{abstract}
\vspace{1cm}
\section{Introduction}
One of the fundamental problems in space physics has been the expansion dynamic of the solar wind \cite{hundhausen2012, dessler1967, parker1969, parker1960}. 
In this context, different models have studied micro and macroscopic physics, from studies of heating and acceleration to turbulence and instabilities \cite{Matteini.2011,cranmer2015,bruno2016turbulence, hellinger2015, Moya.2014}. 
Among these, research has sought to study and understand the role of the expansion in solar wind dynamics. 
One of the most emblematic models, including the expansion in the kinetic and magnetohydrodynamics (MHD) equations, is the double adiabatic theory or Chew, Goldberger, and Low (CGL) \cite{Chew.1956, Hunana_2019}. 
This theory describes the thermodynamic evolution of the plasma starting from a hydrodynamic model without collisions and heat flux. 
The well-known double adiabatic equations can describe the evolution of plasma's temperatures or pressures. 
These equations allow us to correlate the temperature anisotropy ($A = T_\perp/ T_\parallel = \beta_\perp / \beta_\parallel$) and the parallel plasma beta ($\beta_\parallel = 8\pi nk_B T_\parallel/B_0^2$, where $\parallel, \perp$ denote directions with respect to the background magnetic field $\textbf{B}_0$), for a radial decreasing magnetic field and density.
Thus, Matteini et al. \cite{Matteini.2007, Matteini.2011} compared the CGL model with proton's data measured by Helios at various distances from 0.3 to 1 AU from the Sun. These studies showed that the evolution of anisotropy and plasma beta was not well described by the CGL equations, suggesting additional heating processes, which the theory does not predict and still has not been well described \cite{marsch1982wave,Moya.2012, Moya.2014}.

These results, among others, have motivated diverse research works intending to understand the heating and acceleration phenomena in the solar wind \cite{Ofman.2011, barnes1968, barnes1992}.  
First, we must understand and parameterize the physical context to decipher the mechanisms that can affect plasma dynamics. 
One of the fundamental factors that allow us to understand the solar wind's dynamics is the potential role of particle collisions.
Measures have shown that at 1 AU in the solar wind the particle density, temperature, and mean free path are from the order of 10 cm$^{-3}$, 10$^5$ K and 10$^7$ km, respectively \cite{Marsch.2006}.
Therefore, the low density coupled to the higher mean free path of particles implies that Coulomb collisions are not frequent enough to establish a thermodynamic or Maxwellian equilibrium \cite[empirically defined the collisional regime]{Livi.1986}. 
For relaxing to equilibrium, the plasma free energy can be released in the form of waves that interact with particles or excite micro-instabilities. 
Thus, non-linear wave processes, such as wave-particle or wave-wave interactions, might play a fundamental role in the dynamics and heating of the solar wind \cite{tam1999,hollweg2002}. 

In this context, it is fundamental to understand and characterize the role of plasma expansion in the dynamics of instabilities. A simple model, such as CGL theory, predicts that even for an isotropic state with small beta values, the plasma (initially stable) can excite the fire-hose instability only by the radial expansion of the plasma. Therefore, studying these phenomena considering the plasma expansion, might allow us to understand and quantify the expanding effects in the evolution of macro and microscopic quantities and its relation with the excitation of instability thresholds \cite{Innocenti_2019_onset,Innocenti.2020}. 

 To understand the role of the expansion in the heating phenomena \textcite{Velli.1992} proposed an expanding model for the solar wind known as the Expanding Box Model (EBM) \cite{Grappin.1996,Liewer.2001}. The EBM allows us to study plasma expansion in a new reference system, co-moving with the plasma parcel. Through coordinate transformations, it is possible to describe the spherical expansion through a Cartesian description of coordinates in which the volume of the plasma box remains constant. This last property has the computational advantage of optimizing the limited memory in simulations, as the volume is fixed, and we can still include expanding effects into the equations and simulations. To maintain a constant simulation volume, work must be done in the walls to counter the intrinsic expansion of the plasma. This work is translated into non-inertial/fictitious forces that modify the conservative form of several equations (i.e., continuity, momentum, energy, etc.).

This model has motivated different research that has studied expansion's role in plasma heating dynamics. For instance, wide applications can be found in theory and simulations or micro and macroscopic scales \cite{Liewer.2001, Ofman.2011, Innocenti.2019, Innocenti.2020, Micera.2021}. In particular, even though this model was initially proposed for MHD simulations, it has been generalizing, for example, the acceleration regions' dynamics in the expanding solar wind \cite{Tenerani_2017}. More recent research has focused on the characterization of the expanding effects at microscopic scales through an extension of the EBM to a quasi-linear model of the solar wind \cite{Seough_2023}, which couple the double adiabatic equations with quasi-linear theory and the possible effect of the EB parameters on the macroscopic evolution. Nevertheless, this model does not include the EBM coordinate transformations from a first principle approach in which the quasi-linear equations can be modified due to the expanding parameters. On the other hand, \textcite{Echeverria-Veas_2023} developed a first principle description for plasma expansion using the EBM. They studied the EB modifications in the kinetic equation given by the Vlasov formalism, in which it is possible to explore the effects of expansion in the dynamics of microscopic scales. One of the main results of this research is that they obtained the same EBM-MHD equations published in the literature but from a first principles approach. This description also generalizes the pressure equation (in an EBM context) into its tensorial expression, allowing us to explore a more complex system in which a scalar polytropic equation for the pressure is insufficient to describe plasma dynamics.

This work aims to study, for the first time, the consequences of the EBM tensorial pressure equation and the radial evolution of the macroscopic quantities. As we explained before, CGL theory motivated the study of plasma expansion as it can trigger instability thresholds. Nevertheless, the double adiabatic description does not fully incorporate the intrinsic properties of the expansion (e.g., expanding velocity), and only a radial decreasing density and magnetic field profiles are needed. For those reasons, we aim to study how a complete description of plasma expansion based on the EBM can modify the conservation of the double adiabatic equations. This study will allow us to characterize and quantify the expanding effects in the macroscopic quantities and their relation with the heating dynamics of the solar wind.

This paper is organized as follows: In Section \ref{section 2} we introduce the Expanding Box Model formalism and the multi-fluid equations written in the EB frame. In Section \ref{section 3} and Section \ref{section 4} we develop the EBM-CGL equations and an approximated solution based on a solar wind-like plasma, with different magnetic field profiles is done in Section \ref{section 5}. Finally, in Section \ref{section 6}, we summarize and conclude our main results with further applications of the model.

\section{Expanding Box Model \& MHD}
\label{section 2}

\begin{figure}
    \centering
    \includegraphics[width= 9cm, height=6cm]{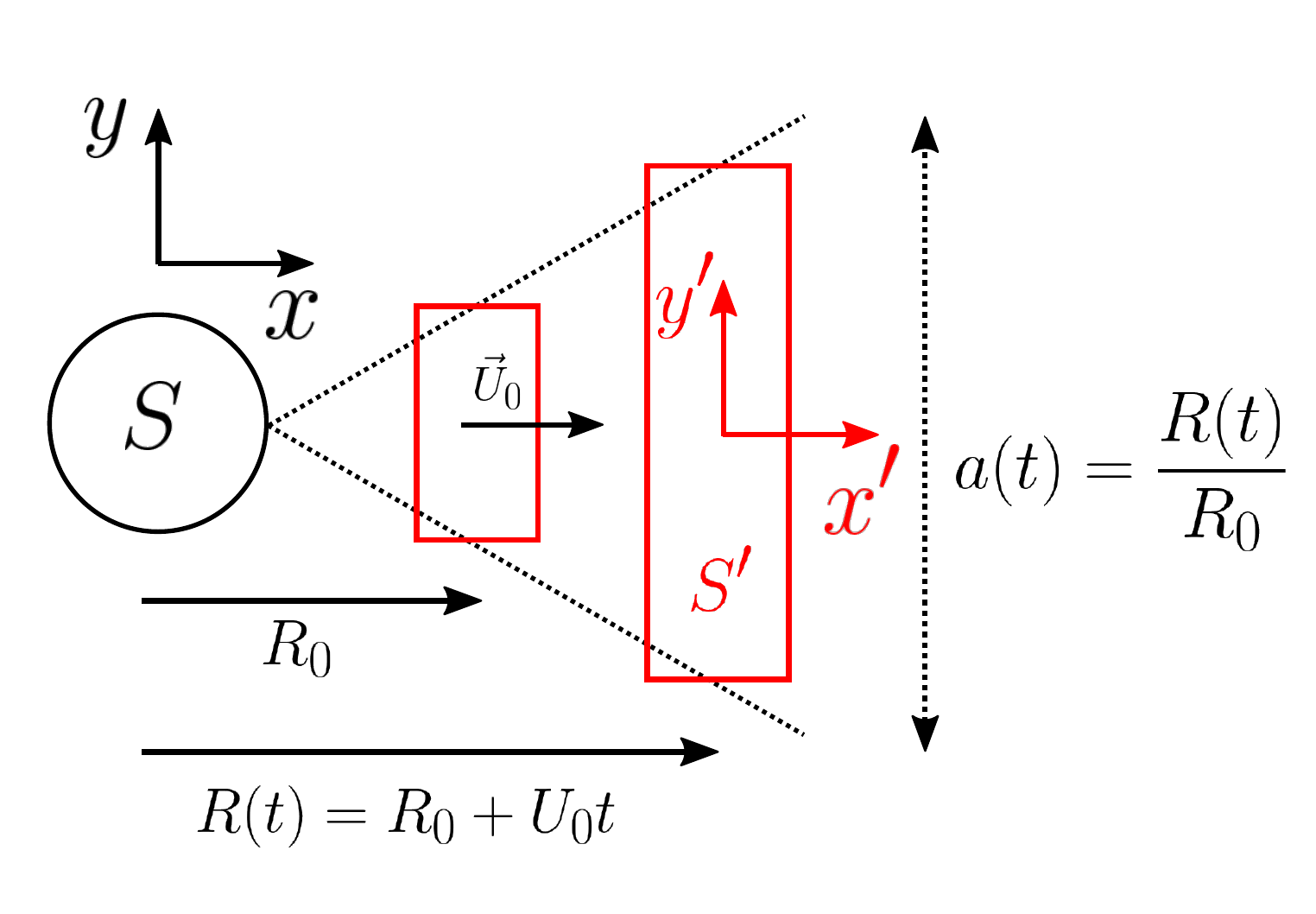}
    \caption{Expanding Box Model Cartesian approximation for a spherically expanding plasma. In red is shown the new system of reference $S'$, co-moving with the plasma parcel. The plasma is initially expanding at distance $R_0$ from the inertial system $S$, at constant velocity $U_0$. In the co-moving frame, the volume of the plasma parcel remains constant.}
    \label{fig: ebm}
\end{figure}

Aiming to develop CGL-like equations, which include the expanding effects of the plasma, we employ the Expanding Box Model (EBM) formalism described by \textcite{Velli.1992, grappin.1993}. The EBM allows us to describe plasma physics in a new system of reference $S'$ co-moving (non-inertial) with the expansion. We can relate the coordinates between the fixed (inertial) $S$ and the $S'$ system through a Galilean transformation in the $\hat{x}$/radial direction and a re-normalization in the perpendicular directions (i.e., $\hat{y}$ or $\hat{z}$)
\begin{align}
    x' &= x - R(t)\,,\label{e2.1}\\
    y' &= \frac{1}{a(t)} y\,,\label{e2.2}\\
    z' &= \frac{1}{a(t)} z \label{e2.3}\,,
\end{align}
where
\begin{align}
    R(t) &= R_0 + U_0 t\,,\label{e2.4}\\
   a(t) &\equiv \frac{R(t) }{ R_0} = 1 + \frac{U_0}{R_0}t\,.\label{e2.5}
\end{align}
In Fig. \ref{fig: ebm} we show the Cartesian approximation and the related quantities to the coordinate transformation. 

Through these coordinate transformations, the electric and magnetic fields change according to a Galilean transformation
\begin{align}
    \textbf{E} = \textbf{E}' - \frac{1}{c} \textbf{U}_0 \times \textbf{B}'\,,\;\;\; \textbf{B} = \textbf{B}'\,. \label{e2.6}
\end{align}

On the other hand, we can establish more relations between both systems for other physical quantities such as spatial gradients, velocity, etc. Following the same notation in \textcite{Echeverria-Veas_2023}, these quantities relate according to
\begin{align}
    \textbf{U}_0 &= U_0 \left( \hat{x} + \frac{y'}{R_0} \hat{y} + \frac{z'}{R_0}\hat{z}\right)\,, \label{e2.7}\\
    \textbf{v} &= \mathbb{A}\cdot \textbf{v}' + \textbf{U}_0 \,, \label{e2.8}\\
    \frac{\partial}{\partial t} &= \frac{\partial }{\partial t'} - \textbf{D}\cdot \nabla'\,, \label{e2.9}\\
    \nabla &=   \mathbb{A}^{-1} \cdot \nabla'\,, \label{e2.10}\\
    \nabla_\textbf{v} &=  \mathbb{A}^{-1}\cdot \nabla_{\textbf{v}'}\,, \label{e2.11}
\end{align}
where
\begin{equation}
        \textbf{D} = U_0 \left(1, \frac{y'}{R}, \frac{z'}{R} \right)\,,
      \quad   \mathbb{A} (t)= \begin{pmatrix}
    1 & 0 & 0 \\
    0 & {a(t)} & 0 \\
    0 & 0 & {a(t)}
    \end{pmatrix}\,.
\end{equation}
Note that $\textbf{v}' =  \left(v_x', v_y', v_z' \right)$ and $\nabla ' = \left(\frac{\partial}{ \partial x'}, \frac{\partial}{\partial y'},\frac{\partial}{\partial z'} \right)$. This notation allows us to work with the same definitions but with primed quantities. All the EB information is outside the definitions through the $\mathbb{A}(t)$ tensor.

Moreover, multi-fluids equations in the co-moving system read as \cite{Echeverria-Veas_2023} (neglecting the primes in the quantities) 
\begin{align}
    &\frac{\partial n}{\partial t} + \nabla\cdot \left( n \textbf{u}\right) = - \frac{2 U_0}{ a R_0} n\,, \label{e2.12}\\
    &\frac{\partial \textbf{u}}{\partial t}  + \left( \textbf{u} \cdot  \nabla\right) \textbf{u} = \frac{q}{m} \left[ \textbf{E} + \frac{1}{c} \left( \mathbb{A} \cdot \textbf{u} \right) \times \textbf{B} \right] \nonumber
   \\ &- \frac{1}{\rho} \nabla \cdot \mathbb{P}  - \frac{U_0}{a R_0} \mathbb{T}\cdot \textbf{u}\,, \label{e2.13} \\
    &\frac{\partial \mathbb{P}}{\partial t} + \nabla \cdot \left( \textbf{u} \mathbb{P} + \mathbb{Q} \right) + \left[ \mathbb{P} \cdot \nabla \textbf{u} \right]^s + \frac{q}{m c} \left[  \textbf{B} \times 
 \left( \mathbb{A} \cdot \mathbb{P} \right) \right]^s \nonumber \\
    &= - \frac{2 U_0}{a R_0} \mathbb{P} - \frac{U_0}{a R_0} \left[ \mathbb{T} \cdot \mathbb{P}\right]^s\,, \label{e2.14}
\end{align}
where the supra-index $s$ represents a symmetric operator that acts on a matrix $\mathbb{C}$ as $\mathbb{C}^s = \mathbb{C} + \mathbb{C}^{T}$ or in index notation $C_{ij} = C_{ij} + C_{ji}$. These equations are also coupled with Maxwell's, nevertheless, as in this paper, we will develop a CGL-like description for expanding plasmas; we will only work with Faraday's Law in the ideal-MHD approximation (also developed in the cited paper)
\begin{align}
    \left( \mathbb{A}^{-1}\cdot \nabla\right)\times\textbf{E} &= -\frac{1}{c}\frac{\partial \textbf{B}}{\partial t} - \frac{U_0}{a R_0 c}\mathbb{L}\cdot \textbf{B}\,, \label{e2.15}\\
    \textbf{E} &= - \frac{1}{c} \left(\mathbb{A} \cdot \textbf{u}\right) \times \textbf{B}\,, \label{e2.16}
\end{align}
where
\begin{align}
\mathbb{L} &= \begin{pmatrix}
         2 & 0 & 0 \\
    0 & 1 & 0 \\
    0 & 0 & 1
    \end{pmatrix}, \quad
     \mathbb{T} = \begin{pmatrix}
    0 & 0 & 0 \\
    0 & 1 & 0 \\
    0 & 0 & 1
    \end{pmatrix}\,.
\end{align}

We recall that these equations are written in the co-moving frame. Note that we have neglected the Hall term in equation \eqref{e2.16}. Continuity \eqref{e2.6} and momentum \eqref{e2.7} equations have been widely studied in EBM contexts; see, for example, references \cite{Grappin.1996, Liewer.2001}.
Pressure equation \eqref{e2.14} is of particular interest as an energetic loss is obtained even if the heat flux is neglected. Note that the right side terms should be related to plasma cooling, as the minus sign represents possible energy losses due to the expansion. This allows us to separate the possible heating/cooling phenomena we are interested in.

The scope of this work is to explore the consequences of plasma expansion in the pressure equation \eqref{e2.8}. In particular, we aim to study how the EBM modifies the double adiabatic invariants described by \textcite{Chew.1956} in CGL theory.

\section{Expanding CGL}
\label{section 3}

In this section, we will focus on developing CGL-like equations from the pressure equation (\ref{e2.14}). Following the same ideas from the CGL equations \cite{Hunana_2018, Webb_2022}, we will also be neglecting the heat flux $\mathbb{Q} = 0$ and only working with the gyrotropic components of the pressure tensor
\begin{align}
    \mathbb{P} = p_{||} \boldsymbol{\tau} \boldsymbol{\tau} + p_\perp \left( \mathbb{I} - \boldsymbol{\tau} \boldsymbol{\tau} \right)\,,\label{e3.1}
\end{align}
where $p_{||}$ and $p_{\perp}$ are gyrotropic components of the pressure tensor, parallel and perpendicular to the magnetic field $\textbf{B}$, respectively; $\boldsymbol{\tau}$ is a unit vector in the direction of the magnetic field
\begin{align}
    \boldsymbol{\tau} = \frac{\textbf{B}}{B}\,. \label{e5.2}
\end{align}

To obtain equations for the evolution of the parallel and perpendicular pressure we will operate with the double contraction in both parallel and perpendicular directions $\boldsymbol{\tau}\boldsymbol{\tau}$ and $\left( \mathbb{I} - \boldsymbol{\tau}\boldsymbol{\tau}\right)$ in equation (\ref{e2.14}) using the definition from equation (\ref{e3.1}), obtaining two equations that describe the evolution for the parallel and perpendicular pressures
\begin{align}
       \frac{1}{p_{||}}\frac{d p_{||}}{d t} - \frac{1}{n} \frac{d n}{d t} + 2 \boldsymbol{\tau} \boldsymbol{\tau} : \nabla \textbf{u} &= - \frac{2 U_0}{a R_0} \left( \tau_y^2 + \tau_z^2 \right)\,, \label{e3.3}\\
    \frac{1}{p_\perp} \frac{d p_\perp}{d t} - \frac{2}{n} \frac{d n}{d t} - \boldsymbol{\tau} \boldsymbol{\tau} : \nabla \textbf{u} &= \frac{U_0}{a R_0} \left( \tau_y^2 + \tau_z^2 \right)\,, \label{e3.4}
\end{align}
where the double contraction between two tensors $\mathbb{A}$ and $\mathbb{B}$ is defined as $\mathbb{A}: \mathbb{B}= A_{ij}B_{ij}\,.$

Left side from Eqs. (\ref{e3.3}) and (\ref{e3.4}) are the usual CGL equations for parallel and perpendicular pressures, which, under these approximations, these equations are equal to zero in the inertial frame $S$. As we can see, expansion effects change the invariance of the equations by adding terms related to the expanding parameter $a$ and the magnetic field components, which mainly affect the transverse directions as expected. For solving these equations, we need to find an expression for $\boldsymbol{\tau} \boldsymbol{\tau}: \nabla \textbf{u}$. As we already used all the previous moments equations (i.e., continuity, momentum, and energy), we will work with Faraday's Law equation \eqref{e2.15} coupled with the condition \eqref{e2.16} and the continuity equation \eqref{e2.12}. Thus, the magnetic field strength in the co-moving frame is given by
\begin{align}
    \frac{1}{B}\frac{d B}{d t} = &\boldsymbol{\tau}\boldsymbol{\tau} : \left[ \left(\mathbb{A}^{-1} \cdot \nabla \right) \left(\mathbb{A} \cdot \textbf{u} \right)\right] + \frac{1}{n}\frac{d n}{d t} \nonumber\\
    &+ \frac{U_0}{a R_0} \left(\tau_y^2 + \tau_z^2 \right) \,. 
\end{align}
Note that the first term on the right side is not exactly $\boldsymbol{\tau}\boldsymbol{\tau} : \nabla \textbf{u}$, as appears in equations \eqref{e3.3} and \eqref{e3.4}. If we write explicitly all the terms it is possible to re-arrange the equation, such that the desired term is obtained
\begin{align}
   \frac{1}{B} \frac{d B}{d t} &=  \boldsymbol{\tau}\boldsymbol{\tau} : \nabla \textbf{u} + \frac{1}{n}\frac{d n}{d t} 
    + \frac{U_0}{a R_0} \left(\tau_y^2 + \tau_z^2 \right)\nonumber\\
    &\;\;\;\;+  \left(a - 1 \right) \left( \tau_x \tau_y \frac {\partial u_y}{\partial x} + \tau_x \tau_z \frac{\partial u_z}{\partial x} \right)\nonumber\\
    & \;\;\;\;+  \left( \frac{1}{a} - 1 \right)\left( \tau_x\tau_y \frac{\partial u_x}{\partial y} + \tau_x \tau_z  \frac{\partial u_x}{\partial z}\right)\,.\label{eq: magnetic field strength EBM}
\end{align}
The terms involving the velocity gradients are the extra terms added and subtracted for obtaining explicitly $\boldsymbol{\tau}\boldsymbol{\tau} : \nabla \textbf{u}$. 

Equations (\ref{e3.3}), (\ref{e3.4}) and (\ref{eq: magnetic field strength EBM}) are the EBM-CGL equations that describe the evolution for parallel and perpendicular pressures when plasma expansion is taken into account from the EBM.
These equations allow us to explore the effects of the expansion in the macroscopic quantities (pressures, plasma beta, and anisotropy). 
 Note that the expansion affects the equations mainly through the expansion parameter $a$ and the transverse components of the magnetic field (i.e., $\tau_y$ and $\tau_z$). 

\subsection{Radial Magnetic Field - Classical CGL }
\label{section 3.1}

For the case when the magnetic field is only in the radial direction $\textbf{B} = B_x \hat{x}$ we recover the classical CGL equations
\begin{align}
     \frac{1}{B} \frac{d B}{d t} =  \boldsymbol{\tau}\boldsymbol{\tau} : \nabla \textbf{u} + \frac{1}{n}\frac{d n}{d t}\,,\\
     \frac{1}{p_{||}}\frac{d p_{||}}{d t} - \frac{1}{n} \frac{d n}{d t} + 2 \boldsymbol{\tau} \boldsymbol{\tau} : \nabla \textbf{u} = 0 \,,\\
    \frac{1}{p_\perp} \frac{d p_\perp}{d t} - \frac{2}{n} \frac{d n}{d t} - \boldsymbol{\tau} \boldsymbol{\tau} : \nabla \textbf{u} = 0\,,
\end{align}
which have by solutions the well-known double adiabatic expressions
\begin{align}
     p_{||} &= \left( \frac{p_{||0} B_0^2}{n_0^3} \right) \frac{n^3}{B^2}  \label{e5.9}\,,\\
    p_\perp &= \left( \frac{p_{\perp 0}}{n_0 B_0}  \right) n B\,. \label{e5.10}
\end{align}
When we assume radial variations for the magnetic field and number density, such as $\sim 1/r^2$, which is suggested by the observations \cite{Maruca_2023}, equations \eqref{e5.9} and \eqref{e5.10} allow us to relate the anisotropy $A = p_{\perp}/p_{||}$ and parallel beta $\beta_{||} = 8 \pi p_{||}/B^2 $ as
\begin{align}
    A = \frac{1}{\beta_{||}}\,. \label{e5.11}
\end{align}
In Appendix \ref{Appendix} is a detailed derivation of this relation. 

The CGL relationship between pressure (or temperature) anisotropy and plasma beta has been widely used when studying solar wind's expansion for protons \cite{marsch1982solar, Marsch.2004, Matteini.2007}. 
Observations have shown a miss-fit when comparing this anticorrelation with measured data between 0.3~AU and beyond 1~AU \cite{Matteini.2007}. 
In an attempt to adjust to the observations for the proton core population in the fast wind,  \textcite{Marsch.2004} have proposed an empirical relation of the form
\begin{align}
    A = \frac{S}{\beta_{||}^b} \label{e5.12}\,,
\end{align}
with $S \simeq 1.16$ and $b\simeq 0.55$. 
On the other hand, CGL equation \eqref{e5.11} predicts a faster decrease of the anisotropy with $b = b_{CGL} = 1$. 
Therefore, there should be alternative mechanisms, e.g.,  heating and cooling, that the double adiabatic approach does not consider. In the next sections, we will focus on the role of the expansion in the EB formalism. 

We recall that the EBM-CGL equations are written in the co-moving system $S'$, and the CGL equations are in the inertial frame, sun-centered $S$. Why do we recover the same expressions with a radial magnetic field in the $x$ direction? As we explained before, the expansion is now affecting mainly the transverse directions (see right side of equations (\ref{e3.3}) and (\ref{e3.4}) for normalized magnetic field components). As in the co-moving system, the box is not expanding in the transverse coordinates, and in the radial or $x$ direction, there is only a Galilean translation; the main effect of the EBM is through the magnetic field. This model does not affect the equations if the magnetic field is only in the radial direction. Therefore, it is expected that the classical CGL equations are recovered for this limit.  

Aiming to explore how the expansion affects relation (\ref{e5.11}), we will solve particular cases for the EBM-CGL equations. Nevertheless, equations are non-trivial as they explicitly depend on the velocity and magnetic field components. In the next section, we will study these equations for different profiles for the physical quantities.

\section{EBM Solution}
\label{section 4}

For the general case when we have all the components of the magnetic field, we can solve pressure equations  (\ref{e3.3}) and (\ref{e3.4}) using Faraday's Law (\ref{eq: magnetic field strength EBM}), which yields to the non-conservative expressions for the CGL adiabatic invariants in the co-moving frame
\begin{align}
    \frac{d }{d t}\left( \frac{p_{||} B^2}{n^3}\right) =  &\frac{2  p_{||} B^2}{n^3} \left[ \left(a - 1 \right) \left( \tau_x \tau_y \frac {\partial u_y}{\partial x} + \tau_x \tau_z \frac{\partial u_z}{\partial x} \right) \right. \nonumber\\
    & \left. +  \left( \frac{1}{a} - 1 \right)\left( \tau_x\tau_y \frac{\partial u_x}{\partial y} + \tau_x \tau_z  \frac{\partial u_x}{\partial z}\right) \right] \,,
\end{align}
\begin{align}
        \frac{d}{d t}\left(\frac{p_\perp}{n B} \right) = & -\frac{p_\perp}{n B}  \left[ \left(a - 1 \right) \left( \tau_x \tau_y \frac {\partial u_y}{\partial x} + \tau_x \tau_z \frac{\partial u_z}{\partial x} \right) \right. \nonumber\\
    & \left. +  \left( \frac{1}{a} - 1 \right)\left( \tau_x\tau_y \frac{\partial u_x}{\partial y} + \tau_x \tau_z  \frac{\partial u_x}{\partial z}\right)\right]\,.
\end{align}
Integrating these equations we can obtain solutions for the parallel and perpendicular pressures
\begin{align}
    p_{||} &= \left( \frac{p_{||0} B_0^2}{n_0^3} \right) \frac{n^3}{B^2} e^{2 I}\,, \label{eq: parallel solution EBM}\\
    p_\perp &= \left( \frac{p_{\perp 0}}{n_0 B_0}  \right) n B e^{-I}\label{eq: perpendicular solution EBM}\,,
\end{align}
where the quantities with the sub-index 0 are the initial conditions at $t=0$ and
\begin{align}
    I = \frac{R_0}{U_0} \int \left\{ \left(a - 1 \right) \left( \tau_x \tau_y \frac{\partial u_y}{\partial x} + \tau_x \tau_z \frac{\partial u_z}{\partial x} \right)\right. \nonumber\\
    \left. +  \left( \frac{1}{a} - 1 \right)\left( \tau_x\tau_y \frac{\partial u_x}{\partial y} + \tau_x \tau_z  \frac{\partial u_x}{\partial z}\right) \right\} d a \,. \label{eq: expanding integral}
\end{align}

Note that when the magnetic field is only radial, we recover the usual CGL equations described at the end of the previous section. On the other hand, we can also recover CGL conditions for non-expanding plasmas as the parameter $a(t) = 1 + U_0\,t/R_0$ is constant for the non-expanding condition $U_0 = 0$. Therefore, the integral $I$ \eqref{eq: expanding integral} vanishes.

Another fundamental and new role appears in the equations. Note that integral \eqref{eq: expanding integral} explicitly depends on the transverse velocity gradients (crossed derivatives), and their consequences in the evolution of macroscopic quantities have yet to be explored. In the following section, we will introduce approximations to solve this integral and the possible physical meaning of these quantities to quantify its effects on the evolution of anisotropy and plasma beta.

We also plotted the results for the anisotropy and parallel beta relations in Fig. \ref{fig2},
\begin{align}
    A = \frac{{p}_\perp}{{p}_{||}}\,, \quad
      \beta_{||} =  \frac{ 8 \pi {p}_{||}}{{B}^2}\,, \label{e4.6}
\end{align}

In the following section, we will focus on solving equations \eqref{eq: parallel solution EBM} - \eqref{e4.6} for two different magnetic field profiles.
\section{Magnetic field profiles}
\label{section 5}

This section will explore two magnetic field profile cases to solve pressure equations \eqref{eq: parallel solution EBM} and \eqref{eq: perpendicular solution EBM}. We can study the anisotropy $A$ and plasma parallel beta $\beta_{||}$ relations and possible deviations from CGL theory by solving these equations. 

Nevertheless, before solving the equations, we will make assumptions and approximations, allowing us to reduce the complexity of the equations. This analysis will give us insights into how the expansion affects the evolution of macroscopic quantities.

To reduce the complexity of the integral in equation \eqref{eq: expanding integral}, we will work with three approximations and assumptions, which are listed below to set the general ideas of the procedure:
\begin{itemize}
    \item Symmetry in the transverse components $y$ and $z$.
    \item Neglecting the radial gradients of the transverse velocity $\frac{\partial u_y}{\partial x}$ and $\frac{\partial u_z}{\partial x}$.
    \item Constant velocity gradients $\frac{\partial u_x}{\partial y} = \frac{\partial u_x}{\partial z} = \eta$ ($\eta$ constant and coupled with the first condition).
\end{itemize} 

Regarding the first point, note that, under the EB transformations, the transverse components $y$ and $z$ are equally affected through the renormalization given by the expanding parameter $a(t) = 1 + U_0\,t/R_0$. Thus, it is expected that if the plasma is only expanding, there is only symmetry in the transverse directions, which differ from radials (i.e., $x$ direction).
From the second point, in the integral (\ref{eq: expanding integral}) of particular relevance are the radial velocity gradients $\frac{\partial u_x}{\partial y}$ or $\frac{\partial u_x}{\partial z}$. 
These terms contain the main variables related to the expansion and the EBM. 
It can show us how the EBM, through the transverse stretching, affects the bulk velocity, quantified by $u_x$.
To explore this idea, as a first approximation, we will only work with these terms in the integral, obtaining
\begin{align}
     I = \frac{R_0}{U_0} \int \left( \frac{1}{a} - 1 \right)\left( \tau_x\tau_y \frac{\partial u_x}{\partial y} + \tau_x \tau_z  \frac{\partial u_x}{\partial z}\right) d a \,. \label{e6.4}
\end{align}

Finally, regarding the third approximation, coupled to the symmetry conditions, the gradients in the integral \eqref{e6.4} are equal $\frac{\partial u_x}{\partial y} = \frac{\partial u_x}{\partial z} $. Nevertheless, we can also consider these gradients constant in the first-order approximation. This last assumption aims to give us an idea of how the integral affects the evolution of physical quantities.

We adopt normalized quantities for computing integral \eqref{eq: expanding integral}. A natural normalization in the integral is the term $R_0/U_0$, which has the exact dimensions as the radial velocity gradients. Therefore, the velocity will be normalized to $U_0$ and the coordinates to $R_0$
\begin{align}
    \Tilde{u}_x = \frac{u_x}{U_0}\,, \quad
    \Tilde{y} = \frac{y}{R_0}\,, \quad \Tilde{z} = \frac{z}{R_0}\,.
\end{align}
With this normalization, the approximations mentioned before read as
\begin{align}
    \frac{U_0}{R_0}\frac{\partial \Tilde{u}_x}{\partial \Tilde{y}} &= \frac{U_0}{R_0} \frac{\partial \Tilde{u}_x}{\partial \Tilde{z}} = \frac{U_0}{R_0} \eta \,,\\
    \tau_y &= \tau_z\,,
\end{align}
where $\eta < 0$ is constant and
\begin{align}
    \eta = \frac{\partial \Tilde{u}_x}{\partial \Tilde{y}} = \frac{\partial \Tilde{u}_x}{\partial \Tilde{z}}\,. \label{e4.12}
\end{align}
Thus, with these normalized quantities and the assumptions made above the integral \eqref{e6.4} is approximated as
\begin{align}
    I_1 \simeq 2 \eta \int\left( \frac{1}{a} - 1 \right) \tau_x \tau_y d a \,. \label{e6.11}
\end{align}

Appendix \ref{Appendix A} is a geometric interpretation of the $\eta$ parameter and its negative value. To solve this integral we only need to set the magnetic field profiles as it depends explicitly on the unitary vector $\boldsymbol{\tau} = \textbf{B}/B$. In the following subsection, we explore the cases of a uniform and non-uniform magnetic field.

\subsection{Uniform magnetic field}

As a first case and approximation to solve the EBM-CGL equations we will work with a radially decreasing density and three-component magnetic field given by
\begin{align}
    n(a) &= \frac{n_0}{a^2}\,,\\
    \textbf{B} (a) &= \frac{B_0}{a^2} \left(1,1,1 \right)\,. \label{eq5.2}
\end{align}

Nevertheless, this magnetic field profile is introduced as a simple mathematical approach in order to set the general ideas of how to solve the equations. A realistic model, which is consistent with the EBM, is developed in the following section. 

From the given magnetic field profile \eqref{eq5.2} the components of the unitary vector $\boldsymbol{\tau}$ are equal
\begin{align}
    \tau_ x = \tau_y = \tau_z = \frac{1}{\sqrt{3}}.
\end{align}
Thus, integral \eqref{e6.11} simplifies to
\begin{align}
    I_1 = \frac{2}{3} \eta \int\left(\frac{1}{a} - 1 \right) d a = \frac{2}{3}\eta \left(\ln (a) - a \right)\,. \label{eq5.10}
\end{align}

From the analytic solution of integral \eqref{eq5.10} we can compute the anisotropy and plasma beta through equations \eqref{eq: parallel solution EBM} and \eqref{eq: perpendicular solution EBM}. Results for this case, under the approximations and considerations developed, are shown in the top panel of Fig. \ref{fig2}.

\subsection{Non-uniform magnetic field}

The first case approximated the problem to give a general idea about how the expansion affects plasma dynamics. Nevertheless, we can study a more consistent model with the EBM equations. We base this case on the results obtained in Particle In Cell (PIC) simulations using the EBM, published by \textcite{Innocenti.2019}. If we have a stretching in the plasma box, magnetic field transverse components are expected to differ from the radials. Therefore, the first case can be improved if we consider these variations.

Therefore, the density and magnetic field profiles are given by \cite{Innocenti.2019}
\begin{align}
    n(a) &= \frac{n_0}{a^2}\,,\\
    \textbf{B} (a) &= \left(\frac{B_{0,x}}{a^2}, \frac{B_{0,y}}{a}, \frac{B_{0,z}}{a}\right)\,. \label{eq50}
\end{align}

These profiles are more consistent with the EB formalism and were obtained directly from the simulations. It is also possible to physically argue these profiles. In Appendix \ref{Appendix B} we developed a first-order expansion of the magnetic field equation in the ideal MHD context, in which the same profile is obtained. 

As we will follow the same ideas developed in the first case, only the fundamental steps will be developed. The magnetic field strength can be written as
\begin{align}
    B = \frac{B_{0,x}}{a^2}\sqrt{1 + 2 \xi^2 a^2 }\,,
\end{align}
where $\xi$ represents the fraction between the magnetic field components at $t=0$
\begin{align}
    \xi = \frac{B_{0,y}}{B_{0,x}} = \frac{B_{0,z}}{B_{0,x}} \,.
\end{align}

Finally, integral \eqref{e6.11} reads as
\begin{align}
    I_2 &=  2 \eta \xi \int \frac{1-a}{1 + 2 \xi^2 a^2 } d a \nonumber\\
    & = \sqrt{2} \eta \arctan\left( 1 + 2 \xi^2 a^2 \right) - \frac{\eta}{2 \xi} \ln\left( 1 + 2 \xi^2 a^2 \right)\,. \label{eq5.16}
\end{align}

\begin{figure*}[h]
    \centering
    \includegraphics[width= 12cm, height= 10 cm]{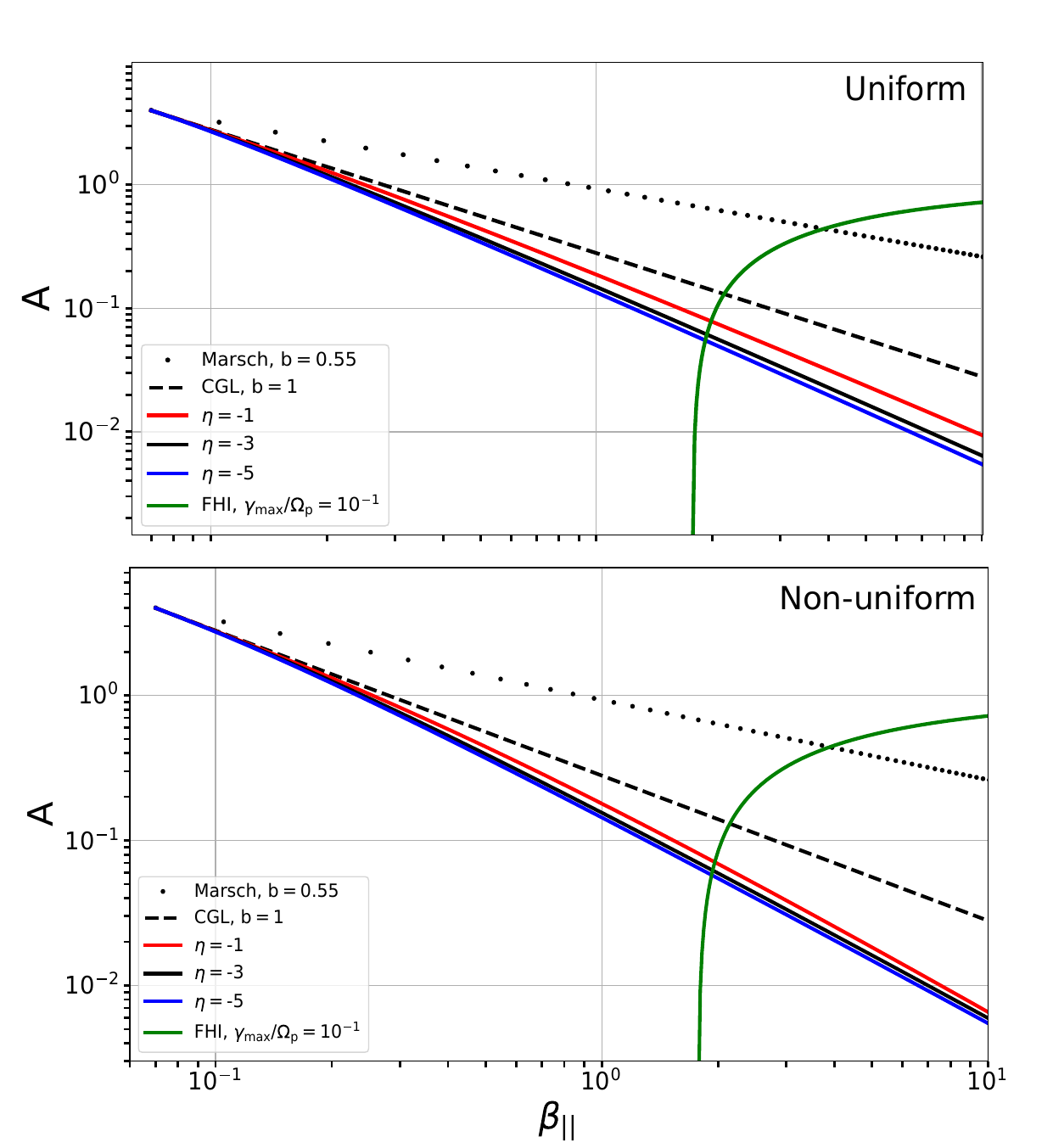}
    \caption{Scatter plot for anisotropy and plasma parallel beta for different values of $\eta = -1, -3, -5$ for the EBM-CGL description (straight lines). Black dashed and dotted lines represent CGL predictions and the empirical relation proposed by Marsch, respectively, given by the anti-correlation $A= S/\beta^{b}_\parallel$. The green line represents the firehose instability (FHI) threshold for a damping rate $\gamma_{max}/\Omega_p = 10^{-1}$. In the top panel are shown the results for the uniform magnetic field (case 1) and in the bottom panel the results for a non-uniform magnetic field (case 2), for $\xi = B_{0,y}/B_{0,x} = 0.1$. For both cases, we set initial values of anisotropy $A = 4$ and plasma beta $\beta_{||} = 0.07$. }
    \label{fig2}
\end{figure*}

Computing integrals \eqref{eq5.10} and 
\eqref{eq5.16} we will be able to study anisotropy and beta relations for the given magnetic field profiles. The non-uniform magnetic field case results are shown in the bottom panel of Fig. \ref{fig2}. 

\begin{figure}[h]
    \centering
    \includegraphics[width= 9cm, height= 6 cm]{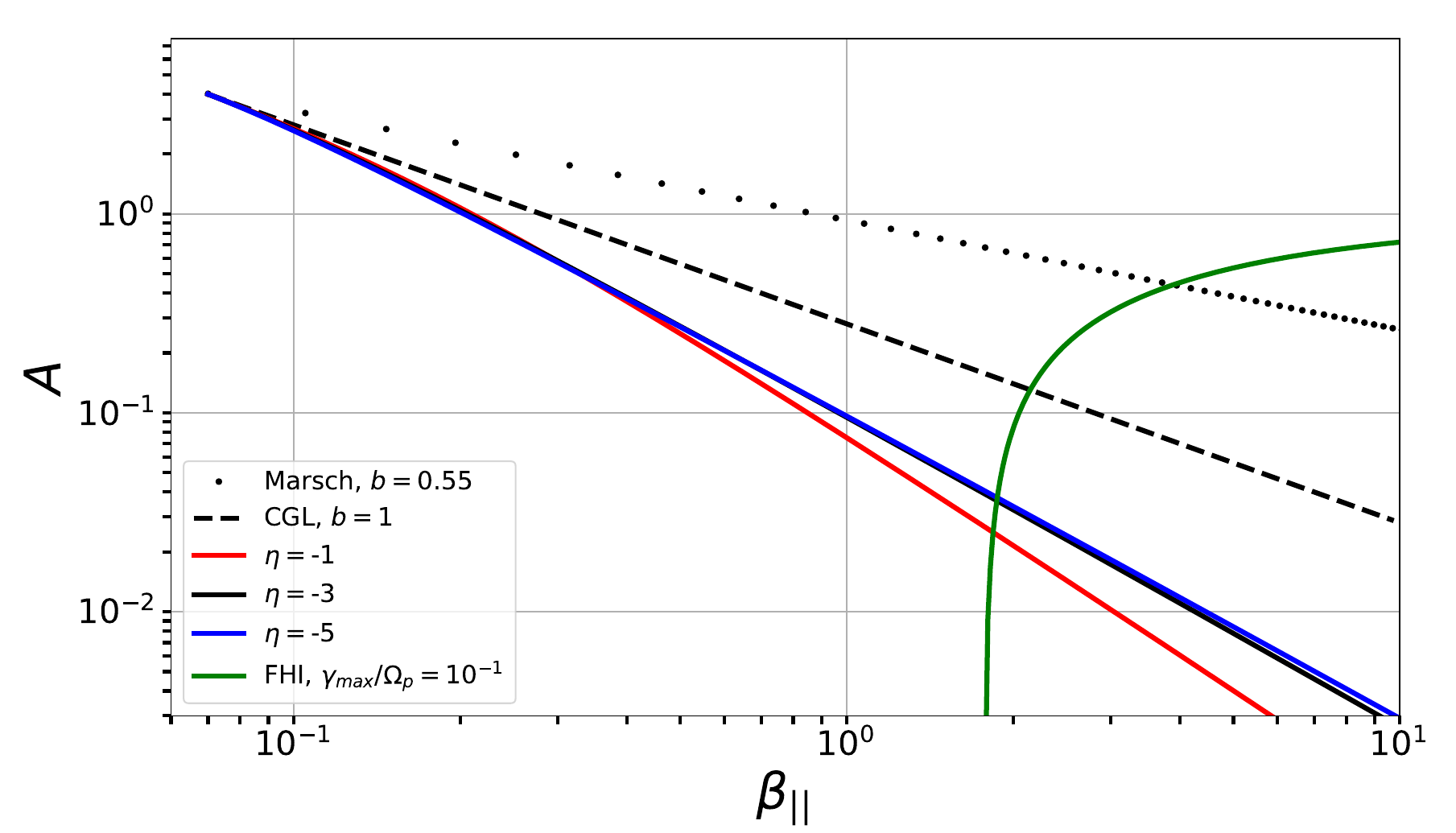}
    \caption{Scatter plot for anisotropy and plasma parallel beta for different values of $\eta = -1, -3, -5$ for the EBM-CGL description (straight lines), for a non-uniform magnetic field with $\xi = B_{0,y}/B_{0,x} = 0.25$. Black dashed and dotted lines represent CGL predictions and the empirical relation proposed by Marsch, respectively, given by the anti-correlation $A= S/\beta^{b}_\parallel$. The green line represents the firehose instability (FHI) threshold for a damping rate $\gamma_{max}/\Omega_p = 10^{-1}$. We set initial values of anisotropy $A = 4$ and plasma beta $\beta_{||} = 0.07$. }
    \label{fig3}
\end{figure}

As in the latter case, we have dependencies on $\eta$ and $\xi$ parameter, we will compute equations for a fixed value of $\xi$ and a variable value of $\eta$. In particular, we will set two values for $\xi$ = 0.1, 0.25, to study from a mainly radial magnetic field ($\xi$ = 0.1) to one with more robust transverse components ($\xi$ = 0.25). The lower value aims to describe a solar wind-like plasma, in which the radial $B_x$ and transverse component $B_y$ vary between -40 to 64 nT and -8 to 8 nT, respectively, on the interplanetary magnetic field (IMF) \cite{King_2005,dessler1967}. Nevertheless, this fraction can increase due to external mechanisms (e.g., geomagnetic storms), and the transverse components might be necessary to describe plasma dynamics fully \cite{Herdiwijaya_2019, Fadiyah_2022}. Therefore, a higher value for $\xi =0.25$ will also be studied in order to set the general behavior of plasma expansion dynamics. In Fig. \ref{fig3} is shown the results for this case.

In Fig. \ref{fig2}, we studied plasma dynamics for the expanding parameter $a$ from initially $a_i=1$ to a final value of $a_f = 10$. We considered initial values of $A_0 = 4$ and $\beta_{||,0} = 0.07$ for anisotropy and parallel beta to mimic observations described in \textcite{Matteini.2007}. Finally, note that integrals \eqref{eq5.10} and \eqref{eq5.16} depend explicitly on the $\eta$ parameter. We plotted results for $\eta = -1,-3, -5$ to study how changes in this parameter affect the evolution of macroscopic quantities; and for the non-uniform case, we plotted results for $\xi = 0.1$. Furthermore, to compare the EBM-CGL solutions with the double adiabatic theory, in Fig. \ref{fig2}, we also plotted the CGL predictions and the empirical relation made by \textcite{Marsch.2004}, adjusted to the evolution of the proton core population for fast wind data, given by equation \eqref{e5.12}. Double adiabatic theory predicts an anti-correlation $A= S/\beta^{b}_\parallel$ with $b=b_\text{CGL}=1$ and the empirical law made by Marsch with $b=b_\text{Marsch} = 0.55$.
We also plotted the firehose instability (FHI) threshold for a damping rate $\gamma_{max}/\Omega_P = 10^{-1}$, (where $\gamma_{max}$ and $\Omega_p$ are the maximum growth rate and proton gyrofrequency, respectively) given by the empirical law \cite{Navarro_2023}
\begin{align}
    \frac{T_\perp}{T_{||}} + \frac{S}{\left( \beta_{||} - \beta_0\right)^c} < 1\,,
\end{align}
where $S= 1.53$, $\beta_0 = 0 $ and $c = 0.74$. It is well known that even for an isotropic state with small values of $\beta_{||}$, the plasma (initially stable) only due to the expansion can excite the FHI. This threshold is shown to establish and compare the values of anisotropy and beta for which the FHI is triggered. We plotted our results for larger values of $\beta_{||}$ to have the general evolution of the macroscopic quantities further from the instability threshold. 

In general, results show a deviation from the adiabatic predictions, and it is visible that the expansion plays a fundamental role as the plasma propagates away from the sun. In particular, the EBM-CGL equations decrease faster than the double adiabatic predictions for both magnetic field cases, further away from Marsch's empirical law. Comparing both magnetic field cases (top and bottom panel in Fig. \ref{fig2}), when a more consistent, non-uniform magnetic field is considered to solve the equations, the anisotropy and beta $A= S/\beta^{b}_\parallel$ relation decreases faster than the uniform case. Even if the results deviate further from CGL predictions and observed data, it is consistent with the idea that plasma dynamics can not be fully adiabatic as the plasma is intrinsically cooling within the expansion. Therefore, if we include these properties through the EBM, we should expect a faster plasma cooling than CGL predictions, deviating from double adiabatic theory. Finally, comparing the non-uniform magnetic field profile for $\xi = 0.1$ (bottom panel in Fig. \ref{fig2}) and $\xi = 0.25$ in Fig. \ref{fig3}, we can notice that as $\xi$ increases anisotropy and beta relations decay faster to the instability threshold. Moreover, it seems to have an inverse role in the $\eta$ parameter as the lowest slope for $\xi = 0.25$ is obtained for $\eta=-1$. We argue that this behavior might be due to the increasing transverse components of the magnetic field. Nevertheless, this result will be explored in future work through a complete analysis of the EBM-CGL equations and possible physical mechanisms that yield this behavior.

The developed equations in this paper allow us to study and quantify the expanding effects of plasma dynamics. Our results suggest that expansion plays a fundamental role in plasma cooling. This behavior allows us to understand the anisotropy and plasma beta evolution by quantifying the expanding effects in the macroscopic quantities. A fundamental role arises from the equation related to the transverse velocity gradients (characterized through the $\eta$ parameter), as its effects have not been fully related to plasma heating/cooling. Nevertheless, the physical interpretation of the $\eta$ parameter was only made by considering the possible geometric variations. To fully describe plasma heating dynamics, a possible relation could be explored between this parameter and kinetic effects, such as wave-particle interactions, which, coupled with plasma expansion, should be fundamental to describe observations.

In this paper, under the related approximations, we have shown that a heating mechanism should be fundamental to counter the plasma cooling due to the expansion. Expansion seems to be exciting the FHI, for lower $\beta_{||}$ values, faster than CGL predictions, which could be a possible energy transfer mechanism. Even though our results deviate further from observations, we can still quantify the expanding effects on the evolution of macroscopic quantities.

\subsection{Inertial and co-moving frame}

We stress that the developed equations in this work are written and solved in the non-inertial frame $S'$, where the primes in the quantities are neglected throughout the article. We recall that, to add the expansion to the equations, we moved the analysis to another system of reference co-moving with the plasma. These transformations are given by a Galilean boost in the radial direction and a renormalization (or stretching) in the transverse direction. At this point, it is natural to wonder about the validity of comparing our results with observational data. This is mainly due to the system of reference in which the results are obtained. For instance, it can be possible to compare the theory and data in different systems of references (sun-centered or co-moving with the solar wind). As the observational details are out of the scope of this paper, we will set the main ideas for switching back to the inertial frame if needed. First, inverting the EBM transformations might lead to the same CGL equations written in the inertial frame. This is because, in the sun-centered (inertial) frame, the equations we started the analysis are the usual ideal MHD and Maxwell's equations without any expanding effects. Therefore, it is expected that if we switch back to the inertial frame, CGL equations will be recovered. 

At this point, it is essential to recall that the EB transformations changed the spatial expansion through temporal variations. Thus, the expansion is no longer a spatial property of the system. Two steps might be needed to switch back to the sun-centered frame and keep the expanding effects in the plasma dynamics. First, the Galilean transformation has to be reversed (i.e., include the constant velocity $U_0$ in the coordinates), and the temporal variations given by the expanding parameter $a(t)$ must be replaced by a radial expansion. 

Moreover, note that the components of the non-uniform magnetic field \eqref{eq50} used in this research are similar to those of a Parker Spiral \cite{Chang_2019}. In particular, both fields vary as $\sim 1/R^2$ and $\sim 1/R$ in the radial and transverse directions, respectively. The EBM replaced these variations through the $a(t)$ parameter. As final remarks, the results we obtain in the co-moving frame are similar to those mentioned, but with temporal variations instead of spatial ones.  Nevertheless, these differences are worth mentioning if switching between systems of references is needed.

\section{Discussions and Conclusions}
\label{section 6}

In this paper, we developed a double adiabatic/CGL-like theory for plasma expansion using the Expanding Box Model (EBM) \cite{Velli.1992,Grappin.1996} and compared it with CGL theory and observations. The EBM allows us to describe plasma physics in a new system of reference co-moving with the expansion by including expanding properties of the plasma into the equations. The theoretical description has the advantage that it is possible to quantify the expanding effects in plasma dynamics by isolating each phenomenon we are interested in.

We started our analysis through the equations developed by \textcite{Echeverria-Veas_2023}, where a multi-fluids description was made in the EB/co-moving frame, obtained from a first principle description given by the Vlasov and Maxwell equations. In this paper, we present, for the first time, a solar wind-like plasma application for the pressure tensor equation developed in the cited paper. Equation \eqref{e2.14} is helpful for the study of more complex systems in which a polytropic equation for the pressure is not enough to describe plasma dynamics.

We studied an ideal-CGL description by working with only the gyrotropic components of the pressure tensor \eqref{e3.1} and neglecting the heat flux from the equations. In Section \ref{section 3}, we developed the EBM-CGL equations from the multi-fluids description and Faraday's Law. Equations \eqref{e3.3}, \eqref{e3.4} and \eqref{eq: magnetic field strength EBM} represent the EBM-CGL equations to solve, and the expanding properties of the plasma (given by the EBM) explicitly modify the conservative form of them, describing the evolution for both perpendicular and parallel pressures. The EBM affects mainly through the magnetic field components, the velocity gradients, and the expanding parameter $a(t) = 1 + U_0\,t/R_0$. It is essential to mention that the EBM-CGL description is not fully adiabatic, as plasma cooling is considered through the EBM rather than the heat flux. This allows us to isolate different phenomena and study their effects on the evolution of plasma expansion. One way to test the consistency of the equations is to study the limit cases where the usual CGL description is recovered. When a purely radial magnetic field is imposed, the equations are reduced to double adiabatic equations. This limit was expected as the EBM affects mainly the transverse components (i.e., $y$ and $z$). The radial direction is not affected by the coordinate transformations (see Section \ref{section 3.1} for the details and the classical CGL equations); therefore, as the EBM-CGL equations are mainly affected by the transverse components of the magnetic field, the classical double adiabatic description was recovered.

On the other hand, in Section \ref{section 4} and \ref{section 5}, we developed our main results in which a solution, under some approximations, for the EBM-CGL equations is obtained. At this point, another limit case naturally appeared from the equations. This second limit is purely related to the expanding velocity as equations \eqref{eq: parallel solution EBM} and \eqref{eq: perpendicular solution EBM} depend explicitly on the evolution of the $a(t)$ parameter. As for a non-expanding plasma $U_0 = 0$, we again recover the classical CGL equations, as expected. To solve the equations fully, a magnetic field profile was needed to compute the radial evolution of parallel and perpendicular pressures. We consider two cases for uniform and non-uniform radially decreasing magnetic fields, allowing us to set the general ideas about how the EBM affects the equations.

To explicitly compare our results with both theory and observations, in Fig. \ref{fig2}, we have presented a scatter plot of the evolution of anisotropy and plasma beta. We plotted the empirical relation proposed by \textcite{Marsch.2004} for the proton core population in fast wind data, which follows a power-law with a decreasing slope of $b_\text{Marsch} = 0.55$. On the other hand, CGL equations predict a slope of $b_\text{CGL} = 1$. Our results deviate further from double adiabatic predictions, and the expansion plays a fundamental role in plasma cooling dynamics as the radial distance increases. 

These results suggest that expansion plays a fundamental role in fully describing observational data for solar wind. As the expansion enhances plasma cooling, we can improve CGL predictions through a more detailed description of the plasma expansion, in which the energetic losses are added through the EBM. We stress that the CGL predictions are only made by neglecting the heat flux and a radially decreasing profile for the density and magnetic field as $\sim 1/a^2$ in an ideal MHD description. These profiles are just a consequence of the expansion, which requires intrinsic properties of the system to describe these dynamics fully. Moreover, the EBM-CGL equations have the advantage as they can include plasma cooling/energetic losses, even when the heat flux is neglected. In this paper, we improved the description of this phenomenon by quantifying the expanding effects in the macroscopic evolution. The main conclusion from these results is that expansion tends to trigger the firehose instability threshold faster than CGL predictions for lower $\beta_{||}$ values. Nevertheless, more than this mechanism is needed to fully understand plasma heating, and possible non-linear effects such as wave-particle or wave-wave interactions also play a fundamental role in plasma heating dynamics. These effects might be explored through the transverse velocity gradients ($\eta$ parameter), which is one of the main reason for the non-conservative form of EBM-CGL equations. 

Further applications to the equations developed in this work can be made. For instance, we recall that the obtained results are under the usual CGL approximations and those related to solving the EBM-CGL equations. In particular, equation \eqref{eq: expanding integral} can be solved for a general velocity gradient profile. We recall that we only considered the radial velocity affected by the transverse gradients related to the EBM. Nevertheless, this velocity profile can be obtained from solar wind data to study how the transverse components are affected by the radial direction and, therefore, its effects on the evolution of the macroscopic quantities. Our work focused on the variables the EBM might affect the plasma expansion dynamics most. On the other hand, we can follow the same ideas developed in CGL theory through the consideration of the Hall term in Faraday's Law, which constitutes the Hall-CGL equations \cite{Hunana_2019}, and explore how the expansion coupled to the Hall term modifies the plasma dynamics. Finally, to explore the possible quasilinear and kinetic effects in the macroscopic quantities, \textcite{Seough_2023} developed an Expanding Box and quasilinear model for the solar wind. They implemented the EB properties of the expansion directly into the classical CGL equations as a global evolution of the quantities. Nevertheless, in this work, we have shown that the EB modification on the CGL equations is not directly due to the effects of the EBM into Faraday's Law, which, coupled with the ideal MHD description, the non-conservative expression for the invariants is not directly to solve. This motivates us to explore the same ideas introduced in their work but with the EBM-CGL equations developed in this paper.

\noindent \textbf{Acknowledgements: } S. Echeverría-Veas is grateful to Agencia Nacional de Investigación y Desarrollo (ANID, Chile) for the National Doctoral Scholarship  N$^\circ$ 21211153. P. S. Moya thanks the support of the Research Vice-rectory of the University of Chile (VID) through grant ENL08/23. M. Lazar and S. Poedts acknowledge support from the Ruhr-University Bochum and the Katholieke Universiteit Leuven. These results were also obtained in the framework of the projects G.0025.23N (FWO-Vlaanderen) and SIDC Data Exploitation (ESA Prodex-12). We also thank Sebastián Saldivia for useful discussions.

\begin{appendix}

\section{Anisotropy and beta CGL relation}
\label{Appendix}

In order to develop the CGL anisotropy and beta relations, we start from the parallel and perpendicular solutions
\begin{align}
     p_{||} &= \left( \frac{p_{||0} B_0^2}{n_0^3} \right) \frac{n^3}{B^2} \,,\\
    p_\perp &= \left( \frac{p_{\perp 0}}{n_0 B_0}  \right) n B\,. 
\end{align}

Assuming radial variations for the density and magnetic field as $\sim 1/r^2$, pressures decreases as
\begin{align}
    p_{||} & \sim \frac{1}{r^2}\,,\\
    p_\perp &\sim \frac{1}{r^4}\,.
\end{align}

Moreover, anisotropy and plasma parallel beta varies as
\begin{align}
    A &= \frac{p_\perp}{p_{||}} \sim \frac{1}{r^2}\,,\\
    \beta & \propto  \frac{p_{||}}{B^2} \sim r^2\,.
\end{align}

Thus, anisotropy and plasma beta, in the CGL approximation, relate according to
\begin{align}
    A = \frac{1}{\beta_{||}}\,.
\end{align}

In \cite{Matteini.2011} there is a detailed derivation and discussions regarding the CGL implications.

\section{Velocity gradients}
\label{Appendix A}

\begin{figure}[h]
    \centering
    \includegraphics[width= 8cm, height=6cm]{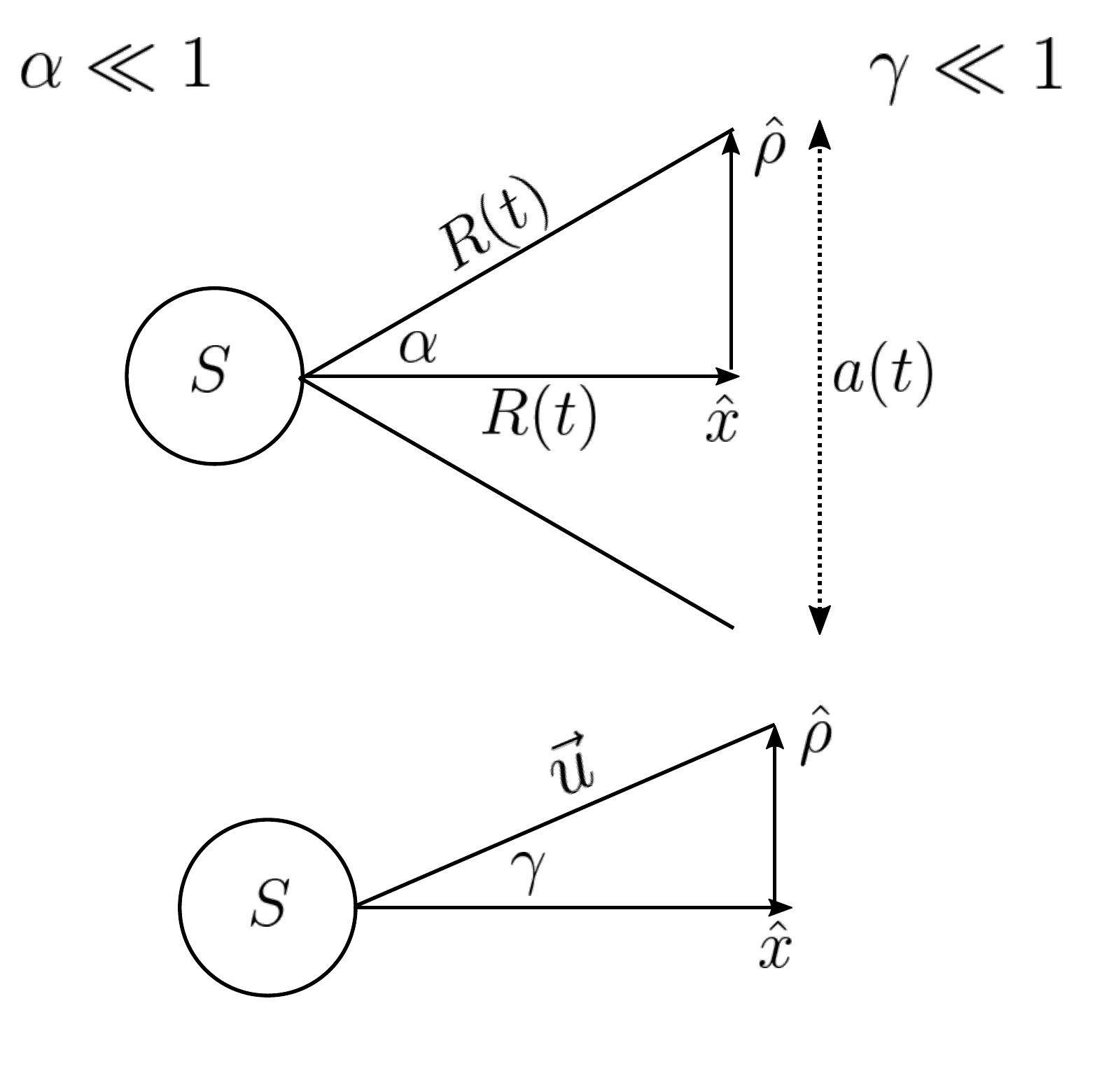}
    \caption{Geometric characterization for the position (top panel) and the velocity (bottom panel). The position of the plasma box is characterized by the angle subtended $\alpha$ between the inertial system $S$ and the box. The plasma velocity within the box is characterized through $\gamma$. 
    Note that the angles $\alpha$ and $\gamma$ can be different and their relation will be given by $\gamma = N \alpha$.}
    \label{fig: ebm_vel}
\end{figure}

We can go further with characterizing the velocity gradients through a two-dimensional analysis. We start our analysis with the position of the plasma box at distance $R(t) = R_0 + U_0 t$, characterized through the subtended angle $\alpha$. Within the box, the bulk's velocity of the plasma might not be characterized with the same angle, as velocity fluctuations might deviate from the constant velocity $U_0$. Thus, we define $\gamma$ as the angle subtended by the direction of expansion ($\hat x$ in inertial frame $S$) and the velocity $\textbf{u}$. We will decompose the velocity into the radial $\hat{x}$ and $\hat{\rho}$ coordinates. As a first approximation, we can relate it with the expanding velocity $U_0$, whose variation will be only due to the trigonometric evolution as Fig. \ref{fig: ebm_vel} shows. We will also consider that the initial transverse direction of the plasma box is small $\alpha \ll 1$ and $\gamma \ll 1$ (see reference \textcite{Liewer.2001} for a detailed discussion of the Cartesian approximation).

Using this geometry a cylindrical description of coordinates fits better for solving the equations. Therefore, the velocity can be written as
\begin{align}
    \textbf{u} &= u_x \hat{x} + u_\rho \hat{\rho} \nonumber\\
    &= U_0 \left(\cos \gamma \hat{x} + \sin \gamma \hat{\rho} \right)\,.
\end{align}
On the other hand, from the geometry is clear that
\begin{align}
    \sin \alpha = \frac{\rho(t)}{R(t)}\,, \quad \cos \alpha = \sqrt{1 - \frac{\rho^2(t)}{R^2(t)}}\,,
\end{align}
which represents the spatial coordinates of the plasma box. Thus, the radial velocity is given by
\begin{align}
    u_x = U_0 \cos \gamma \,.
\end{align}
This expression is useful as the velocity can be related to the polar radius $\rho$ and it is possible to quantify its variation as
\begin{align}
    \frac{\partial u_x}{\partial \rho} &= \frac{d u_x}{d \gamma} \frac{d \gamma}{d \rho} = - U_0 \sin \gamma \frac{d \gamma}{d \alpha} \frac{d \alpha}{d \rho} \nonumber\\
    & = U_0 \left(\frac{\sin \gamma}{\sin \alpha}  \frac{d \gamma}{d \alpha}\right) \frac{d}{d \rho} \left( \cos \alpha \right). \label{e4.19}
\end{align}

Note that expression \eqref{e4.19} relates the velocity gradients between both spatial and velocity angles. We can reduce the free parameters by assuming a relation between the angles, say
\begin{align}
    \gamma = N \alpha\,,
\end{align}
where $N$ $\geq 0$, which represents the fraction between both angles, as in the general case $\alpha$ and $\gamma$ may be different. Finally, assuming small angles $\alpha \ll 1$ and $\gamma \ll 1$, and neglecting second-order terms in $ \rho (t)/R(t)$, equation \eqref{e4.19} yields
\begin{align}
    \frac{\partial u_x}{\partial \rho} \approx - N^2 U_0  \frac{\rho(t)}{R^2(t)}\,, \label{eq4.12a}
\end{align}

Under these approximations, the arch length can be written in terms of the EB variables
\begin{align}
    \rho (t) = \frac{1}{2} R(t) a(t). \label{e4.20}
\end{align}

Finally, normalizing the quantities to the EB variables as
\begin{align}
    \Tilde{u}_x = \frac{u_x}{U_0}\,, \quad \Tilde{\rho} = \frac{\rho}{R_0},
\end{align}
and using equation \eqref{e4.20} in \eqref{eq4.12a} we obtain an explicit value for the $\eta$ parameter
\begin{align}
    \frac{\partial \Tilde{u}_x}{\partial \Tilde{\rho}} = - \frac{N^2}{2} \equiv \eta.
\end{align}
Note that $\rho$ can be both $y$ or $z$ coordinates when comparing with equation \eqref{e4.12}. Due to symmetry in these directions, both gradients are equal, and the same analysis applies. We develop a geometric interpretation of the evolution of the transverse gradients characterized through the $\eta$ parameter. Note that $\eta < 0$ and slight variations of this parameter will be explored to characterize its effects in the evolution of macroscopic quantities. We stress that the transverse gradients in the velocity have not been widely studied in the heating/cooling phenomena of plasma expansion.

\section{Magnetic field radial profile}
\label{Appendix B}

To study the radial variations of the magnetic field components we will start our analysis with the magnetic field equation in an ideal MHD description published by \textcite{Echeverria-Veas_2023}, which also have been studied in previous research \cite[see][for complementary bibliography therein]{grappin.1993, Grappin.1996}. Therefore, the magnetic field equation read as
\begin{align}
    \frac{\partial \textbf{B}}{\partial t} + \left(  \nabla \cdot \textbf{u} \right) \textbf{B} + \left(  \textbf{u} \cdot  \nabla \right) \textbf{B} 
   - &\left[ \textbf{B}\cdot \left(\mathbb{A}^{-1} \cdot \nabla \right) \right]  \left( \mathbb{A}\cdot \textbf{u}\right)\nonumber
   \\& = - \frac{U_0}{a R_0} \mathbb{L} \cdot \textbf{B}\,,
\end{align}
where
\begin{align}
    \mathbb{L} = \begin{pmatrix}
         2 & 0 & 0 \\
    0 & 1 & 0 \\
    0 & 0 & 1
    \end{pmatrix}
    \,,
\end{align}
Through a linear analysis, we will expand the magnetic field into a background $\textbf{B}^{(0)}$ field and a first-order perturbation $\textbf{B}^{(1)}$. In the co-moving frame the velocity only has a first-order perturbation $\textbf{u}^{(1)}$ as this system is already moving along with the plasma parcel with constant velocity $\textbf{U}_0$, therefore, in the $S'$ we will only  work with small fluctuations in the velocity 
\begin{align}
    \textbf{B} &= \textbf{B}^{0} + \epsilon \textbf{B}^{(1)}\,,\\
    \textbf{u} &= \epsilon \textbf{u}^{(1)}\,.
\end{align}
Therefore, the 0th-order equation that the background magnetic fields satisfy, is given by
\begin{align}
    \frac{\partial \textbf{B}^{(0)}}{\partial t} + \frac{U_0}{a R_0} \mathbb{L} \cdot \textbf{B}^{(0)} = 0\,.
\end{align}
From this equation is straight to prove that the magnetic field components are
\begin{align}
    B^{(0)}_x = \frac{B_{0,x}}{a^2}\,, \quad
    B^{(0)}_{y} = \frac{B_{0,y}}{a}\,, \quad B^{(0)}_{z} = \frac{B_{0,z}}{a}\,.
\end{align}
This result is consistent with the simulation results published by \textcite{Innocenti.2019}.
\end{appendix}
\printbibliography

\end{document}